\documentclass[showpacs,10pt,preprintnumbers,amssymb,amsmath]{revtex4}
\usepackage[english,russian]{babel}

\begin{document}

\bibliographystyle{apsrev}

\title{High-Order Correlation Functions \\of Binary Multi-Step
 Markov Chains}

\author{S. S. Apostolov, Z. A. Mayzelis}
\affiliation{V. N. Karazin Kharkov National University, 4 Svoboda
Sq., Kharkov 61077, Ukraine}

\author{O. V. Usatenko\footnote[1]{usatenko@ire.kharkov.ua}, V. A. Yampol'skii
 }
\affiliation{A. Ya. Usikov Institute for Radiophysics and Electronics \\
Ukrainian Academy of Science, 12 Proskura Street, 61085 Kharkov,
Ukraine}

\begin{abstract}

Two approaches to studying the correlation functions of the binary
Markov sequences are considered. The first of them is based on the
study of probability of occurring different ''words'' in the
sequence. The other one uses recurrence relations for correlation
functions. These methods are applied for two important particular
classes of the Markov chains. These classes include the Markov
chains with permutative conditional probability functions and the
additive Markov chains with the small memory functions. The exciting
property of the self-similarity (discovered in Phys. Rev. Lett. 90,
110601 (2003) for the additive Markov chain with the step-wise
memory function) is proved to be the intrinsic property of any
permutative Markov chain. Applicability of the correlation functions
of the additive Markov chains with the small memory functions to
calculating the thermodynamic characteristics of the classical Ising
spin chain with long-range interaction is discussed.

\end{abstract}


\maketitle

\section{Introduction}

The problem of long-range correlated random symbolic systems (LRCS)
has been under study for a long time in many areas of contemporary
physics~\cite{bul,sok,bun,yan,maj,halvin},
biology~\cite{vossDNA,stan,buld,prov,yul,hao},
economics~\cite{stan,mant,zhang},
linguistics~\cite{schen,kant,kokol,ebeling,uya}, etc. Among the ways
to get a correct insight into the nature of correlations of complex
dynamic systems, the use of the \emph{multi-step Markov} chains is
of great importance because it enables constructing a random
sequence with the prescribed correlated properties in the most
natural way~\cite{uyakm,uya,AllMemStepCor}. Also the additive Markov
chains with small memory functions are of interest for the
non-extensive thermodynamics of Ising spin
chains~\cite{equiv,vestnik}.

The multi-step binary Markov chain is characterized by the
\emph{conditional probability} of the defined symbol $a_i$ (for
example, $a_i =1$) occurring \emph{after} previous symbols within
the memory length $N$, i.e., after $N$-word. We define the $L$-word
$T_L$, or the word of length $L$, as the set of $L$ sequential
symbols. The Markov chain can be easily constructed by the
sequential generation of symbols using the prescribed conditional
probability function. The conditional probability points to the
direct association between the symbols. At the same time, the
correlation functions describe the implicit interaction between the
symbols. Correlation functions of all orders determine completely
every statistical property of any random sequence. So, the problem
of finding the correlation functions of high orders is essential.
The aim of the paper is to find the correlation functions of
different orders of the binary Markov chain using two approaches and
to examine their properties.

The paper is organized as follows. The second Section describes
the method of calculating statistical characteristics of the
Markov chain via the conditional probability function. Here the
property of the self-similarity of the correlated random sequences
is also discussed. The third Section is devoted to a method for
finding the correlation functions using recurrence relations for
them. The applications of the proposed general algorithms to some
concrete classes of chains are presented.

\section{ \label{direct} Direct method for calculating
statistical characteristics of the Markov chain}

In this section, we describe a direct method of finding following
characteristics of the $N$-steps Markov chains: the probability
$P(T_N)$ of the $N$-words occurring, the probability $P(T_L)$ of
occurring the words of an arbitrary length $L$, and the high-order
correlation functions.

It is known that all statistical properties of Markov sequence are
determined completely by the conditional probability function. Along
with this function, it is convenient to employ the probabilities
$P(T_N)$ of the $N$-words occurring. All other characteristics of
the chain can be found using these probabilities. Though the
calculation of $P(T_N)$ is quite a challenge in the general case, we
succeed in obtaining the analytical results for some particular
cases (see Subsec.~\ref{permute}).

The \textit{$N$-step Markov chain} is a sequence of random
symbols~$a_i$, $i\in \mathbb{Z}=\{\dots,-2,-1,0,1,2,\dots\}$. It
possesses the following property: the probability of symbol~$a_i$
to have a certain value, under the condition that the values of
all \emph{previous} symbols are given, depends on the values of
$N$ previous symbols only,
\begin{equation}\label{def_mark}
 P(a_i=a|\ldots,a_{i-2},a_{i-1})=
 P(a_i=a|a_{i-N},\ldots,a_{i-2},a_{i-1}).
\end{equation}
We refer to $N$ as \emph{the memory depth}.

In this paper we consider binary chains only, but some results can
be generalized to non-binary sequences as well.

\subsection{Probabilities of the $N$-words occurring}

We start examining the statistical properties of the Markov chains
with searching the probabilities of the $N$-words occurring. There
are $2^N$ different $N$-words in the arbitrary Markov chain which is
characterized by $2^N$ probabilities of these words occurring. They
can be found using evident formulas of the probability theory,
\begin{gather}\label{sum_prob}
P(A)=P(A\cap B)+P(A\cap \overline{B}),
\\\label{cond_prob} P(A\cap B)=P(A|B)P(B).
\end{gather}
Here $A\cap B$ means that events $A$ and $B$ occur simultaneously
and event $\overline{B}$ is opposite to the event $B$.
Eqs.~\eqref{sum_prob},~\eqref{cond_prob} yield the following
equation,
\begin{equation}\label{for_Nword}
P(a_1a_2\ldots a_N)=\sum_{a=0,1}P(a_N|a_{}a_1\ldots
a_{N-1})P(a_{}a_1\ldots a_{N-1}).
\end{equation}
This equation, being written for all possible values of symbols
$a_1,\ldots,a_N$, along with the normalization requirement,
\begin{equation}\label{normal}
\sum_{a_i=0,1 \atop {i=1,\ldots,N}}P(a_1a_2\ldots a_N)=1,
\end{equation}
is in fact the set of linear equations. From this set, one can
obtain all sought probabilities~$P(T_N)$. Though
set~(\ref{for_Nword}), (\ref{normal}) contains $2^N+1$ equations, it
has a single solution because one of Eqs.~\eqref{for_Nword} is a
linear combination of others.

\subsection{\label{cor&word}
Probabilities of arbitrary words occurring and correlation
functions}

Using Eq.~\eqref{sum_prob} we can calculate the probability $P(T_L)$
with $L<N$ by reducing the $N$-words to the $L$-word $(a_1a_2\ldots
a_s)$,
\begin{equation}\label{word_lessN}
P(a_1a_2\ldots a_L)= \sum_{a_i=0,1\atop {i=L+1,\ldots,N}}
P(a_1a_2\ldots a_N),\qquad L<N.
\end{equation}

This equation represents the probability of the $L$-word occurring
as the average of the probabilities to have the $N$-word
containing this $L$-word. It is possible to reduce the $N$-word to
the shorter $L$-word by summing over different sets of symbols at
the left and right edges of the $N$-word.

The word of length $L$ greater than $N$ can be presented as the
combination of the word of length $N$ and $(L-N)$ symbols
following after it. According to the definition of the Markov
chain, each of these $(L-N)$ symbols can take on a certain value
with the probability depending on the precedent $N$-word. The
$(L-N)$-fold use of Eq.~\eqref{cond_prob} yields the following
equation,
\begin{equation}\label{word_moreN}
P(a_1a_2\ldots a_{L})=P(a_1a_2\ldots a_{N}) \prod_{r=1}^{L-N}
P(a_{N+r}|a_{r}a_{r+1}\ldots a_{N+r-1}),\qquad L>N.
\end{equation}

Let us define a correlation function of the $s$th order:
\begin{equation}\label{def_cor1}
K_{s}(i_1,i_2,\ldots,i_{s})=\overline{
(a_{i_1}-\bar{a}_{i_1})(a_{i_2}-\bar{a}_{i_2})\ldots
(a_{i_s}-\bar{a}_{i_s})},
\end{equation}
where $\overline{\cdots}$ is the statistical average over the
ensemble of chains. We consider ergodic chain. According to the
Markov theorem (see, e.g., Ref.~\cite{shir}), this property is valid
for the homogenous Markov chains if the conditional probability does
not take on values $0$ and $1$. In this case averaging over the
ensemble of chains and over the chain coincide.

Formally, function $K_s$ depends on $s$ arguments ($s$ different
indexes of the symbols), but we do consider homogenous Markov
chains. Therefore, correlation function $K_{s}$ depends on $(s-1)$
arguments, i.e., the distances between the indexes,
$r_1=i_2-i_1,r_2=i_3-i_2, \ldots, r_{s-1}=i_s-i_{s-1}$. Its
definition is written as
\begin{equation}\label{def_cor1}
K_{s}(r_1,r_2,\ldots,r_{s-1})=\overline{
(a_0-\bar{a})(a_{r_1}-\bar{a})(a_{r_1+r_2}-\bar{a})\ldots
(a_{r_1+\ldots+r_{s-1}}-\bar{a})}.
\end{equation}
Here $\bar{a}$ is the average number of unities in the sequence and
notation $\overline{\cdots}$ is the statistical average over the
chain,
\begin{equation}\label{epsilon-av}
\overline{f(a_{r_1},\ldots ,
a_{r_1+\ldots+r_{s}})}=\lim_{M\to\infty}\frac{1}{2M+1}
\sum_{i=-M}^{M}f(a_{i+r_1},\ldots , a_{i+r_1+\ldots+r_{s}}).
\end{equation}

Introducing the notation,
\begin{equation}
R_0=0,\qquad R_k=\sum\limits_{i=1}^k r_i, \qquad
d_{i}=a_i-\bar{a},
\end{equation}
we rewrite definition~\eqref{def_cor1} of the correlation function,
\begin{equation}
K_{s}(r_1,r_2,\ldots,r_{s-1})=\sum_{a\!_{R_i}=0,1\atop
{i=0,\ldots,s-1}}
 d_{R_0}d_{R_1}\ldots
d_{R_{s-1}} P(a_{R_0}a_{R_1}\ldots a_{R_{s-1}}).
\end{equation}

Now we can complement the set of symbols $a_{R_0},a_{R_1},\ldots,
a_{R_{s-1}}$ with the symbols between them. Finally we have the
following formula,
\begin{equation} \label{cor1}
K_{s}(r_1,r_2,\ldots,r_{s-1})=\sum_{a_i=0,1\atop
{i=0,\ldots,R_{s-1}}}
 d_{R_0}d_{R_1}\ldots
d_{R_{s-1}} P(a_{0}a_{1}\ldots a_{R_{s-1}}).
\end{equation}
Here probabilities $P(a_{0}a_{1}\ldots a_{R_{s-1}})$ should be
calculated using Eqs.~\eqref{word_lessN} or~\eqref{word_moreN}.

\subsection{Permutative Markov chains} \label{permute}

Solving linear system~(\ref{for_Nword}),~\eqref{normal} for the
general case is a challenging problem. Here we demonstrate the
application of the described method to a certain class of the
Markov chains. We assume that the conditional probability function
of the chain under consideration is independent of the order of
symbols in the previous $N$-word. We refer to such sequences as
the \emph{permutative} Markov chains.

\subsubsection{Probabilities of word occurring in the
permutative Markov chain} \label{permute_word}

It is convenient to introduce a new abbreviated notation for the
conditional probability function $P(a_{N+1}|a_1a_2\ldots
a_N)=p_k(a_{N+1})$. Here $k=a_1+a_2+\ldots+a_N$ is the number of
unities in $N$-word $(a_1a_2\ldots a_N)$. Besides, we define
$p_k(1)$ as $p_k$.

Now we seek the solution of
system~(\ref{for_Nword}),~\eqref{normal} in the form
\begin{equation}\label{word_per}
b_N(k)=b_N(a_1+a_2+\ldots+ a_N)=P(a_1a_2\ldots a_N).
\end{equation}
In other words, the probability of the $N$-word occurring depends
on the number of unities in this $N$-word only. Then
Eq.~\eqref{for_Nword} can be rewritten as the following recurrence
relation,
\begin{equation}
b_N(k)=p_{k-1}b_N(k-1)+p_kb_N(k).
\end{equation}

The solution of this equation is
\begin{equation}\label{for_Nw_per}
b_N(k)=b_{N}(0)\prod_{r=1}^{k}\dfrac{p_{r-1}}{1-p_r}.
\end{equation}
Here probability~$b_{N}(0)$ can be obtained from
Eq.~\eqref{normal},
\begin{equation}\label{for_Nw0_per}
b_{N}(0)=\Big(\sum\limits_{k=0}^NC_N^{k}
\prod\limits_{r=1}^{k}\dfrac{p_{r-1}}{1-p_r}\Big)^{-1},
\end{equation}
\[ \quad C_n^k=\frac{n!}{k! (n-k)!}=\frac{\Gamma (n+1)}{\Gamma (k+1) \Gamma (n-k+1)}.
\]

The probability $P(a_1a_2\ldots a_N)$, Eq.~\eqref{word_per}, does
not depend on the order of symbols in $N$-word $(a_1a_2\dots
a_N)$. From Eq.~\eqref{word_lessN} and the above statement it
follows that the probability $P(T_L)$ of a short word (with $L<N$)
occurring is likewise independent of the order of symbols in this
word. Denoting this probability by
$b_L(k)=b_L(a_1+a_2+\ldots+a_L)=P(a_1a_2\ldots a_L)$ we arrived at
the following formula,
\begin{equation}\label{w_lessN_per}
b_L(k)=\sum_{m=0}^{N-L}C_{N-L}^mb_N(k+m).
\end{equation}

The probability of the long $L$-word (with $L>N$) occurring
\textbf{does depend} on the order of the symbols in this word.
Using Eq.~\eqref{word_moreN} we have
\begin{equation}\label{w_moreN_per}
P(a_1a_2\ldots
a_{L})=b_N(q_0)\prod_{r=1}^{L-N}p_{q_{r-1}}(a_{N+r}).
\end{equation}
Here $q_i=a_{i+1}+a_{i+2}+\ldots+ a_{{i+N}}$.
Equations~(\ref{for_Nw_per})-(\ref{w_lessN_per}) are the
generalization of results earlier obtained in
\cite{Biased,AllMemStepCor} for the additive binary Markov chain
with the step-wise memory function.

\subsubsection{Correlation functions} \label{permute_cor}

Here we present analytical results from the calculation of the
correlation functions of arguments $r_1,\ldots,r_{s-1}$ satisfying
the condition $r_1+\ldots+r_{s-1}<N$. For this purpose we express
the fraction of unities in the chain using
Eq.~\eqref{w_lessN_per},
\begin{equation}
\overline{a}=b_1(1)=\sum_{k=1}^Nb_N(k)C_{N-1}^{k-1}.
\end{equation}
 For the permutative Markov chains under consideration, the correlation
function of arguments $r_1,\ldots,r_{s-1}$ depends on their number
$s$ only. Equation~\eqref{cor1} yields
\begin{equation}
K_{s}(r_1,r_2,\ldots,r_{s-1})=K_{s} =
\sum_{k=0}^Nb_N(k)S_N(k,s,\overline{a}),
\end{equation}
\begin{equation}
S_N(k,s,\bar{a})=\sum_{j=0}^{\min\{s,k\}}
(-\bar{a})^{s-j}C_{N-j}^{k-j}C_{s}^j.
\end{equation}
In particular, the binary correlation function $K_2(r)$ of any
permutative chain is constant for the values of arguments less
than the memory depth: $\quad K_2(r)=K_2, \quad r<N$.

If we apply all derived formulas to the additive Markov chain with
the step-wise memory function,
\begin{equation}
p_{k} =\frac{1}{2}-\nu+\mu \left(\frac{2k}{N}-1\right),
\end{equation}
we get the
 correlation functions of order $s$,
\begin{equation}
K_{s}=\frac{\Gamma(n_1+n_2)}{\Gamma(n_1)}\sum_{k=0}^{s}
(-\bar{a})^{s-k}C_{s}^k\frac{\Gamma(n_1+k)}{\Gamma(n_1+n_2+k)}.
\end{equation}
 Here
\begin{equation}
\bar{a}=\frac{n_1}{n_1+n_2},\quad n_1=
\frac{N(1-2(\mu+\nu))}{4\mu},\quad n_2=
\frac{N(1-2(\mu-\nu))}{4\mu}.
\end{equation}
For $s=2$ we recover the results previously obtained in \cite
{Biased,AllMemStepCor} for the binary correlation function $K_2$.

\subsection{Self-similarity of the permutative Markov chain}

In this subsection we point out an interesting property of the
permutative Markov chains, namely, their self-similarity. The
discussion of this issue for the step-like memory function is
presented in paper~\cite{Biased}.

Let us reduce the $N$-step Markov sequence by regularly (or
randomly) removing some symbols and introduce the decimation
parameter $\lambda=N^*/N,\,1/N <\lambda < 1$, which represents the
fraction of symbols kept in the chain.

Due to approachment of symbols in the sequence after the decimation
procedure, the binary correlation function $K_2(r)$ transforms into
\begin{equation}
\label{K_2^*} K_2^*(r)=\begin{cases}
(1-\{r/\lambda\})K_2([r/\lambda])+\{r/\lambda\}K_2([r/\lambda]+1),&
\mbox{regular decimation,}\\
\sum\limits_{l=0}^\infty K_2(r+l) C_{l+r-1}^l(1-\lambda)^l
\lambda^r,&
\mbox{random decimation,}\\
\end{cases}
\end{equation}
where $[x]$ is the maximal integer number less than $x$ and
$\{x\}=x-[x]$.

The correlation function $K_2(r)$ of the permutative Markov chain
is equal to constant value $K_2$ with arguments $r\leqslant N-1$.
According to Eq.~\eqref{K_2^*} (regular decimation), the values of
the correlation function $K_2^*(r)$ at $r\leqslant \lambda(N-1)$
take on the same value $K_2$. In the case of the random decimation
we have only an exponentially small (at $N\gg1$) difference
between the correlation function $K_2^*(r)$ and value $K_2$,
\begin{equation} |K_2^*(r)-K_2|\leqslant
\frac{A}{\sqrt{\bar{\lambda}N}}
\exp\big(-a(\bar{\lambda},\lambda)N\big), \qquad
r=\bar{\lambda}N,\qquad \bar{\lambda}<\lambda.
\end{equation}
Here the new function $a(\bar{\lambda},\lambda)$ and constant $A$
are introduced,
\begin{equation}
a(\bar{\lambda},\lambda)
=\bar{\lambda}\ln\left(\dfrac{\bar{\lambda}}{\lambda}\right)
+(1-\bar{\lambda})\ln\left(\dfrac{1-\bar{\lambda}}{1-\lambda}\right)>0,
\qquad A=\frac{e}{\sqrt{\pi}}\big(|K_2|+\max_{r\geqslant
N}|K_2(r)|\big).
\end{equation}

Thus, the correlation function of the decimated sequence is
constant (or asymptotically constant) at the region $r<N^*$. This
property is referred to as the \textit{self-similarity}.

The self-similarity is the intrinsic property of the permutative
Markov chains. It is possible to show that, in the general case,
the conditional probability function of the decimated $N$-step
Markov sequence is of infinite memory depth. This is because its
conditional probability is ''blurred'' by the decimation procedure
and becomes dependent on \textbf{all} previous symbols.

So, the self-similarity is the property of the binary correlation
function only and inheres in all random sequences with a constant
binary correlation function $K_2(r)$ at $r<N$,
\begin{equation}\label{self} K_2(r)=
K_2,\, 1\leqslant r < N \Rightarrow \mbox{Self-similarity.}
\end{equation}

\section{Correlation functions and characteristic equations}

In this Section, we deal with the recurrence relations to find the
correlation functions. In Subsection A we show that these relations
yield the explicit expression for the correlation functions via the
roots of the characteristic equations. Subsection B is devoted to
applications of this method. The last Subsection contains some
generalizations of the above-mentioned recurrence relations.

The procedure of finding the correlation functions $K_s$ is based on
the mathematical induction method. In other words, we suppose that
all correlation functions $K_l$ of the orders less than $s$ are
found. For the convenience sake, we admit that $K_0=1$ and $K_1=0$.

\subsection{
\label{cor_for_add} High-order correlation functions of the additive
Markov chain}

Consider the $N$-step Markov chain with the \emph{additive}
conditional probability function,
\begin{equation}\label{def_add_m}
P(a_i=1|a_{i-N}\dots a_{i-1}) =\bar{a}+\sum \limits
_{r=1}^{N}F(r)(a_{i-r}-\bar{a}).
\end{equation}
Here function $F(r)$, $r=1,\ldots,N$, is referred to as the
\textit{memory function} and $\bar{a}$ is the fraction of unities in
the above sequence (for details see Ref. \cite{mel}).

Let us find the recurrence relations for the correlation functions
of $N$-step additive Markov chain. For this purpose, we first
calculate explicitly the average over symbol
$a_{r_1+\ldots+r_{s-1}}$ in Eq.~\eqref{def_cor1}. Using the notation
of Sec.~\ref{cor&word} and Eq.~\eqref{def_add_m}, and taking into
account equation $P(a_{R_{s-1}}=1|\cdot)+ P(a_{R_{s-1}}=0|\cdot)=1$,
we can rewrite Eq.~\eqref{def_cor1} for arbitrary $r_i>0$,
$i=1,\ldots,s-1$, in the form,
\[
\overline{(a_{R_0}-\bar{a})\ldots (a_{R_{s-1}}-\bar{a})}=\]
\begin{equation}
=\overline{(a_0-\bar{a})\ldots (a_{R_{s-2}}-\bar{a})((1-\bar{a})
P(a_{R_{s-1}}=1|T_{N,R_{s-1}})- \bar{a}
P(a_{R_{s-1}}=0|T_{N,R_{s-1}}))}=
\end{equation}
\[= \overline{(a_0-\bar{a})\ldots
(a_{R_{s-2}}-\bar{a})\sum \limits
_{r=1}^{N}F(r)(a_{R_{s-1}-r}-\bar{a})}.\] Here $T_{N,R_{s-1}}$ is
the set of symbols $(a_{R_{s-1}-N},\ldots,a_{R_{s-1}-1})$. In that
way, we obtain the fundamental recurrence relation connecting the
correlation functions of different orders~$s$,
\begin{equation}\label{recur_cor}
K_s(r_1,\dots,r_{s-1})=\sum \limits
_{r=1}^{N}F(r)K_s(r_1,\dots,r_{s-1}-r).
\end{equation}

In the particular case $s=2$, this equation,
\begin{equation}\label{recur_cor2}
K_2(r_1)=\sum \limits _{r=1}^{N}F(r)K_2(r_1-r),
\end{equation}
was obtained and discussed in \cite{mel}. Recurrence
relation~\eqref{recur_cor} is correct for $r_i>0$, $i=1,\ldots,s-1$.
Provided that $r_{s-1}\leqslant N$, the last argument of the
correlation function in the right-hand side of Eq.~\eqref{recur_cor}
is negative or zero and one should interpret it in the following
manner, which is referred to as \textit{''collating''}. If the
correlation function has negative arguments, we must reorganize it
according to definition~\eqref{def_cor1}. For example,
\[K_4(2,2,-3)=\big\langle
(a_{0}-\bar{a}) (a_{2}-\bar{a})(a_{4}-\bar{a})
(a_{1}-\bar{a})\big\rangle=\]
\begin{equation}
=\big\langle (a_{0}-\bar{a})(a_{1}-\bar{a})
(a_{2}-\bar{a})(a_{4}-\bar{a}) \big\rangle=K_4(1,1,2).
\end{equation}
 If the correlation function has zero arguments (indexes $i$ and $k$ of two
multipliers $(a_{i}-\bar{a})$ and $(a_{k}-\bar{a})$ coincide) one
should employ the property of the \emph{binary} chain: $a_i^2=a_i$,
$a_i=\{0,1\}$. Thus, we have
 \[(a_i-\bar{a})^2=(1-2\bar{a})(a_i-\bar{a})+
 \bar{a}(1-\bar{a}).\]

With this property we can write the useful relations for the
correlation function containing zero arguments in different
positions among all arguments of $K_s$,
\begin{gather}
\label{start_zero}
\begin{split}K_s(0, r_2, \dots, r_{s-1}) = (1-2\bar{a})
K_{s-1}(r_2,\dots,r_{s-1})+ \bar{a}(1-\bar{a})K_{s-2}(r_3,\dots,
r_{s-1}),\\K_s(r_1,\dots, r_{k-1}, 0, r_{k+1}, \dots, r_{s-1}) =
(1-2\bar{a})
K_{s-1}(r_1,\dots, r_{k-1}, r_{k+1}, \dots, r_{s-1})+\\
+\bar{a}(1-\bar{a}) K_{s-2}(r_1,\dots, r_{k-1}+r_{k+1},\dots,
r_s),\quad k\neq 1,s, \\
K_s(r_1,\dots, r_{s-2}, 0) = (1-2\bar{a})
K_{s-1}(r_1,\dots,r_{s-2})+ \bar{a}(1-\bar{a})K_{s-2}(r_1,\dots,
r_{s-3}).
\end{split}
\end{gather}

The general solution of Eq.~\eqref{recur_cor} can be represented as
a linear combination,
\begin{equation}\label{sol_rec}
K_s(r_1,\dots, r_{s-1})=\sum\limits_{j=1}^{N}L_j(r_1, \dots,
r_{s-2}) \xi_j^{r_{s-1}},\,r_i>0\,(i=1,\ldots,s-1),
\end{equation}
of powers of the roots $\xi_j$, $j=1,\ldots,N$, of the
characteristic equation,
\begin{equation}\label{char_eq}
\xi^N-\sum\limits_{j=1}^N F(j)\xi^{N-j}=0.
\end{equation}
New functions~$L_j(r_1, \dots, r_{s-2})$, i.e. coefficients of
linear form  Eq.~\eqref{sol_rec}, in their turn, should be defined.
All one has to do is to substitute Eq.~\eqref{sol_rec} into
''collated'' Eq.~\eqref{recur_cor} for $0<r_{s-1}<N$. Finally, this
procedure yields the recurrence relations for the sought functions
$L_j$.

The general solution of this recurrence relation is reduced to the
form,
\begin{equation}\label{solrsmo}
L_i(r_1, \dots, r_{s-2})=\sum\limits_{k=1}^{N(N-1)/2}
M_{ik}(r_1,\dots, r_{s-3})\eta_k^{r_{s-2}},
\end{equation}
where $\eta_k$, $k=1,\ldots,N(N-1)/2$, are the roots of the new
characteristic equation
\begin{equation}
\det \Upsilon(\eta)=0,
\end{equation}
\begin{equation}
\Upsilon_{ij}(\eta)=(\eta/\xi_j)^{i-1}-\xi_j^{i-1}+\delta_{1i},
\quad i,j=1,\ldots,N.
\end{equation}
From the ''collating'' procedure we can find the next recurrence
relations for functions $M_{ij}$ etc. Using this algorithm we can
find, in principle, the correlation functions of all orders.

\subsection{Application of the algorithm}

All results obtained in this subsection are correct for the additive
Markov chain, but some of them are valid for the non-additive chain
as well.

The first simple result is that the correlation functions of all odd
orders are zero for the additive Markov chain with $\bar{a}=1/2$.
This results from Eq.~\eqref{start_zero}: function $K_{2m+1}(\cdot)$
is expressed only in term of $K_{2m-1}(\cdot)$ and $K_1=0$.

The correlation function of the second order $K_{2}(\cdot)$ (the
\emph{binary} correlation function) for an additive Markov chain can
be found from Eq.~\eqref{recur_cor}:
\begin{equation}
K_2(r)=\sum_{j=1}^NL_j\xi_j^r,\quad r>-N,
\end{equation}
where $\xi_j$, $j=1,\ldots,N$, are the roots of the characteristic
equation~\eqref{char_eq}. Coefficients $L_j$ should be obtained by
the collating:
\begin{equation}
K_2(0)=\bar{a}(1-\bar{a});\quad K_2(-r)=K_2(r),\quad 0<r<N.
\end{equation}

Paper~\cite{AllMemStepCor} contains an analysis of this equations in
the case of the additive Markov chain with the step-wise memory
function. Below we present the results of calculation for the
correlation function of a weakly correlated chain,
\begin{equation}
|P(a_i=1|a_{i-N}\ldots a_{i-1})-\bar{a}|\ll 1.
\end{equation}
 This case is very important from the physical point of view because the statistical
properties of the equilibrium long-range correlated Ising chains can
be represented and considered as the Markov chains. Specifically,
the Markov chain with the small memory function is statistically
equivalent to the weakly correlated Ising chain (see
Refs.~\cite{equiv,vestnik}).

For the correlation function of even order of the unbiased
($\bar{a}=1/2$) additive chain we have
\begin{gather}\label{cor_2_weak}
K_2(r)\approx \dfrac{1}{4}F(r),\quad 0<r<N,\quad
K(0)=\dfrac{1}{4},
\\ \label{cor_2s_weak}
K_{2s}(r_1,\dots,
r_{2s-1})=\prod\limits_{j=1}^{s}K_2(r_{2j-1}),\quad
r_j>0,\,j=1,\ldots,2s-1.
\end{gather}
The latter equation is correct for the non-additive sequence as
well. In the case of one-step Markov chain, $N=1$, this equation is
exact. But if $N>1$ one should interpret it as the asymptotical
equality. At $N=2$, the \emph{exact} result for the correlation
function $K_4(\cdot)$ of the additive chain is,
\begin{equation}
 K_4(r_1,r_2,r_3)=K_2(r_1)K_2(r_3)+
(-\xi_1\xi_2)^{r_2}\left(\dfrac{1}{4}K_2(r_1+r_3)-K_2(r_1)K_2(r_3)\right).
\end{equation}
Here $\xi_1$ and $\xi_2$ are the roots of the characteristic
equation,
\begin{equation}
\xi^2-F(1)\xi-F(2)=0.
\end{equation}

And finally, Eq.~\eqref{recur_cor} allows us to express the memory
function in terms the prescribed correlation functions $K_s$. In
Ref.~\cite{mel}, the method of constructing the correlated binary
sequence with prescribed binary correlation function $K_2$ was
earlier discussed.

\subsection{Generalized algorithm of finding the correlation functions}

Here we generalize the algorithm proposed in
Subsec.~\ref{cor_for_add} for the additive Markov chains to the
arbitrary binary multi-step Markov chains.

The conditional probability function of the $N$-step binary Markov
chain can be written as
\begin{equation}\label{def_gen_m}
P(a_i=1|a_{i-N}\ldots a_{i-1})=\sum_{l_j=0,1\atop {j=1,\ldots,N}}
F(l_1,l_2,\ldots, l_N)\prod_{r=1}^N(a_{i-r}-\bar{a})^{l_r}.
\end{equation}
It is a general form for the arbitrary binary function, because
$P(\cdot|\cdot)$ can be thought of a linear function of each
argument $a_j$. We refer to function $F(l_1,l_2,\ldots, l_N)$ as the
\textit{generalized} memory function.

Equation~\eqref{def_gen_m} can be applied to the additive Markov
chain described by Eq.~\eqref{def_add_m}, namely,
\begin{gather}
F(0,0,\ldots,0)=\bar{a}, \notag\\
\quad F(\underbrace{0,\ldots,0}_{r-1},1,\underbrace{0,\ldots,
0}_{N-r})=F(r),\quad r=1,2,\ldots,N,\\
F(l_1,l_2,\ldots, l_N)=0,\quad l_1+l_2+\ldots+l_N>1.\notag
\end{gather}

Now we obtain the recurrence relation for the correlation function
with the coefficients expressed via the generalized memory function.
 according to the procedure performed in Subsection~\ref{cor_for_add}, we substitute the last symbol
$a_{r_1+\ldots+r_{s-1}}$ in Eq.~\eqref{def_cor1} (for $r_j>0$,
$j=1,\ldots,s-1$) for its conditional probability $P(\cdot|\cdot)$
in the form of Eq.~\eqref{def_gen_m}. As a result one gets,
\begin{gather}\label{gen_rec_cor}
K_s(r_1,\ldots,r_{s-1})=
(F(0,0,\ldots,0)-\bar{a})K_{s-1}(r_1,\ldots,r_{s-2})+
\\ \notag+\sum_{\substack{l_j=0,1\\ j=1,\ldots,N}}
F(l_1,l_2,\ldots, l_N)
K_{s+k-1}(r_1,\ldots,r_{s-2},r_{s-1}-\rho_{k},
\rho_{k}-\rho_{k-1},\ldots,\rho_2-\rho_{1}),
\end{gather}
 where $k=l_1+\ldots+l_N\not=0$. The increasing numbers $\rho_j$,
 $j=1,\ldots,k$ are the indexes, for which $l_{\rho_j}\neq 0$.

Note that some summands in Eq.~\eqref{gen_rec_cor} can have
non-positive arguments if $r_{s-1}\leqslant N$. For this case, one
should apply the ''collating'' procedure to this equation.

Thus, the algorithm for finding of the correlation functions can be
formulated as follows:
\begin{enumerate}
    \item To obtain the correlation function
    $K_s(r_1,r_2,\ldots,r_{s-1})$ at $r_i>0$, $i=1,2,\ldots,s-1$,
    we should use Eq.~\eqref{gen_rec_cor} and execute the
    ''collating'' procedure. We find that function
    $K_s(r_1,\ldots,r_{s-1})$ is expressed via the correlation
    functions with the sum of their arguments less than initial
    sum $r_1+\ldots+r_{s-1}$.

    \item  Using Eq.~\eqref{gen_rec_cor} and performing the
    ''collating'' procedure $m$ times with respect
    to the obtained correlation functions, we derive a
    relation between function $K_s(r_1,\ldots,r_{s-1})$ and some
    other correlation functions.
    \begin{enumerate}
     \item If all of them have the order less than $s$, we obtain
     the
     recurrence relation for the correlation function of the
     $s$th order. This relation can be solved by the method of
     characteristic equations without executing item~\ref{item}.
     (A similar case takes place for the additive Markov chain.)

     \item In the opposite case, for $m>r_1+r_2+\ldots+r_{s-1}-N$,
     function $K_s(r_1,\ldots,r_{s-1})$ is expressed via
     the correlation functions with
     the sums of their arguments less than $N$. Some of these
     functions are of the order greater than $s$. Then we should go
     to the next item of the algorithm.
    \end{enumerate}

    \item \label{item} All correlation functions with the sum
    of arguments less than $N$ should be found from a set of
    linear equations. To this end we write
    Eq.~\eqref{gen_rec_cor}
    for every such correlation function. After executing the
    ''collating'' procedure we obtain the set of
    $\big(2^{N-1}-1\big)$ linear equations for
    $\big(2^{N-1}-1\big)$ sought values of the correlation
    functions.
\end{enumerate}

In some particular cases, it is convenient to define $K_s(\cdot)$ in
Eq.~\eqref{def_cor1} and $P(\cdot|\cdot)$ in Eq.~\eqref{def_gen_m}
using an arbitrary fixed value $\tilde{a}$ (for example,
$\tilde{a}=1/2$) instead of the average $\bar{a}$. In this incident
every reasoning remain valid provided that the value of $\bar{a}$ is
changed to $\tilde{a}$ in Eqs.~\eqref{gen_rec_cor} and
\eqref{start_zero}, and $K_1=0$ is changed to
$K_1=\tilde{a}-\bar{a}$. To obtain average $\bar{a}$ we should add
Eq.~\eqref{gen_rec_cor}, written for $K_1$, to the set of equations
in item~\ref{item} of the algorithm. Then we arrive at the set of
$2^{N-1}$ linear equations and $\big(2^{N-1}-1\big)$ sought values
of the correlation functions and one of the average $\bar{a}$.
Besides, one can use $\tilde{a}$ instead of $\bar{a}$ in the
different physical problems, which are relevant to the Markov chain.
For instance, the energy in the Ising model is more convenient to
express in terms or function
$K_2(r)=\overline{(a_{i}-1/2)(a_{i+r}-1/2)}$.

\section{Conclusion}

Thus, we have demonstrated two approaches to determining the
statistical properties of the binary Markov chains. The first of
them should be used if the probabilities of $N$-words occurring can
be easily found. The correlation functions of different orders can
be expressed via these probabilities. The examples of these chains
is the permutative Markov sequences. The second approach allows one
to find the correlation functions directly from the recurrence
relations connecting them with the memory function. In the general
case, these relations contain the correlation functions of different
orders and, hence, they are difficult to solve. In the case of
additive chains, this relations are simplified and their use helps
to find the solutions.

\end{document}